\documentclass[11pt]{article}

\usepackage{amsmath}
\usepackage{amsfonts}
\usepackage{amsthm}
\usepackage{amssymb}
\usepackage{fullpage}
\usepackage{dsfont}
\usepackage{hyperref}
\usepackage{graphicx}

\newtheorem{theorem}{Theorem}
\newtheorem{lemma}[theorem]{Lemma}

\theoremstyle{definition}
\newtheorem{defn}{Definition}

\newcommand{\superscript}[1]{\ensuremath{^\textrm{#1}}}

\newcommand{\abs}[1]{\left\lvert #1 \right\rvert}

\newcommand{\complex}{{\mathbb C}}
\newcommand{\reals}{{\mathbb R}}

\newcommand{\tensor}{\otimes}

\newcommand{\adjoint}{\dagger}

\newcommand{\ket}[1]{|#1\rangle}
\newcommand{\bra}[1]{\langle #1|}
\newcommand{\ketbra}[2]{\ket{#1}\!\bra{#2}}        
\newcommand{\density}[1]{\ketbra{#1}{#1}}     

\newcommand{\set}[1]{{\left\{#1\right\}}}    

\newcommand{\norm}[1]{\left\|\,#1\,\right\|}       
\newcommand{\enorm}[1]{\norm{#1}_{\mathrm{2}}}      
\newcommand{\fnorm}[1]{\norm{#1}_{\mathrm {F}}}    

\newcommand{\trace}{{\rm Tr}}

\newcommand{\kp}{\ket{\psi}}
\newcommand{\rhocc}{\rho_{\operatorname{cc}}}
\newcommand{\rhow}{\rho_W}
\newcommand{\rhobe}{\rho_{BE}}
\newcommand{\rhoent}{\rho_{\rm {ent}}}
\newcommand{\fu}{\operatorname{d}(\rho,U^B)}

\newcommand{\fum}{\operatorname{d}_{\operatorname{max}}(\rho)}

\newcommand{\fucc}{\operatorname{d}_{\operatorname{max}}(\rhocc)}

\newcommand{\fuw}{\operatorname{d}_{\operatorname{max}}(\rhow)}
\newcommand{\fube}{\operatorname{d}_{\operatorname{max}}(\rhobe)}
\newcommand{\fua}{\operatorname{d}(\rho_a,U^B)}
\newcommand{\fual}{\operatorname{d}(\rho_\alpha,U^B)}
\newcommand{\fuam}{\operatorname{d}_{\operatorname{max}}(\rho_a)}
\newcommand{\fualm}{\operatorname{d}_{\operatorname{max}}(\rho_\alpha)}

\begin{document}
\thispagestyle{empty}


\title{
On Global Effects Caused by Locally Noneffective Unitary Operations}
\author{Sevag Gharibian \footnote{School of Computer Science
and Institute for Quantum Computing, University of Waterloo, Waterloo, Canada}
\and Hermann Kampermann \footnote{Institut f\"{u}r Theoretische Physik III,
Heinrich-Heine-Universit\"{a}t D\"{u}sseldorf, D\"{u}sseldorf, Germany}
\and Dagmar Bru\ss\ \footnotemark[2]}

\maketitle \thispagestyle{empty}

\begin{abstract}
Given a bipartite quantum state $\rho$ with subsystems $A$ and $B$ of
arbitrary dimensions, we study the entanglement detecting capabilities
of locally noneffective, or cyclic, unitary operations
[L.~B.~Fu, Europhys. Lett., vol. 75, pp. 1--7, 2006]. Local cyclic
 unitaries have the special property that they leave
their target subsystem invariant. We investigate the  distance between
$\rho$ and the global state after local application of such unitaries as a
possible indicator of entanglement. To this end, we derive and discuss
closed formulae for the maximal such distance
 achievable for three cases of interest: (pseudo)pure quantum states,
Werner states, and two-qubit states. What makes this criterion interesting,
as we show here, is that it surprisingly displays behavior similar to recent
 anomalies observed for non-locality measures in higher dimensions, as well
 as demonstrates an equivalence to the CHSH inequality for certain classes
 of two-qubit states. Yet, despite these similarities, the criterion is not itself a non-locality measure. We
also consider entanglement detection in bound entangled states.
\end{abstract}

\section{Introduction}\label{scn:introduction}

Over the last two decades, quantum entanglement has been the subject of
 intense research, due to the continuing discoveries of interesting uses
for the phenomenon by the quantum computing and information community
(see~\cite{B02} and~\cite{HHH07} for surveys). One of the  remaining open
 problems, however, is that of deciding separability of a quantum state -
 that is, given a (in our case, bipartite) state $\rho$ acting on
the Hilbert space
$\mathcal{H}^M\otimes\mathcal{H}^N$, where $M$ and $N$ denote the
respective dimensions of the subsystems, decide whether $\rho$ can be written
 in the form
\begin{equation}\label{defn:separable}
    \rho=\sum_{k=1}^n p_k\ketbra{a^k}{a^k}\tensor\ketbra{b^k}{b^k},
\end{equation}
for $p_k\in\reals^+$, $\sum_{k=1}^{n}p_k=1$, $n\geq 1$, $\ket{a^k}\in
\mathcal{H}^M$, $\ket{b^k}\in\mathcal{H}^N$, and $\enorm{a^k}=\enorm{b^k}=1$,
 for $\enorm{\cdot}$ denoting the Euclidean norm. A state which can be written in this form is called \emph{separable}, and if additionally we have $n>1$, we refer to the state as \emph{classically correlated}. This problem was proven
 NP-hard by Gurvits~\cite{G03}, implying that
it is highly unlikely for a general
 solution to exist for all possible inputs $\rho$. Another topic of interest
 which has recently garnered renewed attention is the study of entanglement
 and non-locality as comprising distinct resources~\cite{E93,ADGL02,BGS05,MV08}---for generalized Bell inequalities for qutrits, or the ability to close the detection loophole, for example, it has been found that non-maximally entangled quantum states outperform maximally entangled states. Here, we study an approach to detecting entanglement which surprisingly also displays effects similar to those seen for such entanglement versus non-locality arguments.

The approach we consider was originally proposed by Fu~\cite{F06}.
It consists of applying a unitary operation to one of the subsystems,
while demanding that the density matrix of the
subsystem is invariant under this transformation.
However, the global density matrix may be changed,
and therefore there may be a non-zero distance between the original (global) state and the one after applying the local unitary operation. We will call this quantity ``Fu
distance'', and will be particularly interested in its maximal value, for a given initial state,
where the maximisation is over all locally noneffective unitaries.

Note that the principle underlying this approach has long been implicitly harnessed, for example, in superdense coding~\cite{BW92}, where applying a Pauli operator to half of a Bell state can give rise to an orthogonal Bell state.

This paper is organized as follows. In Section~\ref{scn:preliminaries}, we
define the Fu distance and
discuss its relevant properties. In Sections~\ref{sscn:pseudopure}
and~\ref{sscn:werner} we derive closed formulae for the maximal Fu distance
 for pseudopure states (i.e. mixtures of a projector onto a pure state with
the identity) and Werner states, respectively. In
Section~\ref{scn:CHSHconnections}, we discuss
 connections between the CHSH inequality and the maximal Fu distance
 for two-qubit systems. In Section~\ref{scn:be}
we present the attempt to detect bound entanglement using the Fu distance.
 Finally, we conclude in
Section~\ref{scn:conclusions} and pose open questions.

\section{Definition and properties of the Fu Distance}\label{scn:preliminaries}

 Given a quantum state $\rho$, acting on
$\mathcal{H}^{M}\tensor\mathcal{H}^N$, with
density matrices of the
subsystems $\rho_A=\trace_B(\rho)$ and $\rho_B=\trace_A(\rho)$,
 define a locally noneffective, or \emph{cyclic}~\cite{F06}, unitary operation $U^B$, as one satisfying the condition $U^B\rho_B{U^B}^\adjoint=\rho_B$. This is equivalent to demanding
\begin{equation}
    [\rho_B,U^B]=0.\label{eqn:cyclicdef}
\end{equation}
Then, letting $\rho_f=(I\tensor U^B)\rho(I\tensor{U^B}^\adjoint)$, our quantity of interest is
\begin{equation} \label{defn:fudistance}
    \fu:=\frac{1}{\sqrt{2}}\fnorm{\rho-\rho_f},
\end{equation}
which we dub the Fu distance~\cite{F06}, and where
$\fnorm{A}=\sqrt{\trace(A^\adjoint A})$
denotes the Frobenius norm (or Euclidean norm)
for matrices. Thus, we are applying a local unitary operation which leaves the target reduced state invariant, and yet may produce a global shift in the joint system, the quantification of which we will
study as a possible indicator of entanglement. To this end, we will be most interested in the quantity
\begin{equation}
    \fum := \max_{\text{cyclic }U^B}\fu ,
\end{equation}
i.e. the maximal possible global distance achieved under any locally noneffective
unitary operation.

Let us briefly discuss some relevant properties of $\fum$. First, note that
Equation~(\ref{defn:fudistance}) can be straightforwardly rewritten in the useful form~\cite{F06}
\begin{equation}\label{eqn:fudistancerewritten}
    d(\rho,U^B)=\sqrt{\trace(\rho^2)-\trace(\rho\rho_f)},
\end{equation}
from which it is easy to see that $0\leq \fu \leq 1$, with the latter inequality saturated if and only if $\rho$ is pure and orthogonal to $\rho_f$. For any product state, i.e. $\rho=\rho_A\otimes\rho_B$, it is clear that $\fum=0$~\cite{F06}. It is not known whether $\fum>0$ for all entangled states, although we will later show that this is in indeed the case for all entangled pseudopure and Werner states.

One can find an upper bound for the Fu distance
of any \emph{classically correlated} state $\rhocc$ in a bipartite system, by
generalizing an argument of Fu~\cite{F06} to dimensions $M$ and $N$ for the subsystems:
\begin{equation}\label{eqn:ccupperbound}
    \fucc\leq \sqrt{\frac{2(M-1)(N-1)}{MN}}.
\end{equation}
The derivation of  Equation (\ref{eqn:ccupperbound}) is as follows:
 Any bipartite state $\rho$ can be written in Fano form~\cite{F83}:
\begin{equation}
    \rho = \frac{1}{MN}\left(I^A\otimes I^B + \vec{r}^A\cdot\vec{\sigma}^A\otimes{I^B}+
                I^A\otimes\vec{r}^B\cdot\vec{\sigma}^B+\sum_{i=0}^{M^2-1}\sum_{j=0}^{N^2-1}T_{ij}\sigma^A_i\otimes\sigma^B_j\right),\label{eqn:fano}
\end{equation}
where $I$ denotes the identity matrix, $\vec{r}^A$ denotes the $(M^2-1)$-dimensional Bloch vector for subsystem $A$ with $r^A_i=\frac{M}{2}\trace(\sigma^A_i\rho_A)$, $\vec{\sigma}^A$ denotes the $(M^2-1)$-component vector of traceless Hermitian generators for
$SU(M)$, and the matrix $T$ is a real matrix known as the \emph{correlation matrix}, whose entries are given by $T_{ij}=\frac{MN}{4}\trace(\sigma^A_i\otimes\sigma^B_j\rho)$. The definitions for subsystem $B$ are analogous.

Equation~(\ref{eqn:fudistancerewritten})
can now be rewritten via straightforward manipulation as~\cite{F06}
\begin{equation}
    d(\rho,U^B)=\frac{2}{MN}\sqrt{\sum_{ij}T_{ij}^2-\sum_{ij}T_{ij}T_{ij}^f},\label{eqn:fuT}
\end{equation}
where $T^f$ is the correlation matrix for $\rho_f$. Let us derive bounds on each sum in the square root. First, for a separable state
$\rhocc=\sum_{k=1}^{n}p_k\ketbra{a^k}{a^k}\tensor\ketbra{b^k}{b^k}$, one has $T_{ij}=\sum_kp_kr^{A_k}_ir^{B_k}_j$, where $\vec{r}^{A_k}$ and $\vec{r}^{B_k}$ are the Bloch vectors corresponding to states $\ketbra{a^k}{a^k}$ and $\ketbra{b^k}{b^k}$, respectively~\cite{F06}. Via the Cauchy-Schwarz inequality, we thus have
\begin{equation}
    \sum_{ij}T_{ij}^2=\sum_{ij}\left(\sum_kp_kr^{A_k}_ir^{B_k}\right)^2\leq\sum_kp_k\left(\sum_ir^{A_k}_i\right)\left(\sum_jr^{B_k}_j\right)\leq\frac{MN(M-1)(N-1)}{4},
\end{equation}
where the last inequality follows from the fact that $\enorm{r^{A_k}}\leq \sqrt{M(M-1)/2}$ (resp. for $\enorm{r^{B_k}}$)~\cite{Kim03}. A lower bound of $\sum_{ij}T_{ij}T^f_{ij}\geq-\sum_{ij}T^2_{ij}$ is also easily found using the Cauchy-Schwarz inequality. Substituting these into Equation~(\ref{eqn:fuT})
 gives Equation~(\ref{eqn:ccupperbound}), as desired.

Note that the bound in  Equation (\ref{eqn:ccupperbound}) is only non-trivial
for small dimensions $M$ and $N$, as $\fu\leq 1$ has to hold.
For $M=N=2$, Equation (\ref{eqn:ccupperbound}) gives a tight bound\footnote{Consider, for example, $\rho_{cc}=\frac{1}{2}\ketbra{00}{00} + \frac{1}{2}\ketbra{11}{11}$ and $U^B=\sigma_x$, the Pauli X operator.} of $\fucc\leq 1/\sqrt{2}$, and for $M=N=3$, a possibly loose bound of $\fucc\leq \sqrt{8/9}$.
Determining a tight upper bound on $\fucc$ for arbitrary dimensions remains at present an intriguing open problem.

\section{Maximizing the Fu Distance}\label{scn:maxfudist}

\subsection{Pseudopure States}\label{sscn:pseudopure}
We now derive a closed formula for $\fum$ for pseudopure states, and follow with a discussion of its implications. Specifically, consider a pseudopure quantum state $\rho$ acting on
$\mathcal{H}^{M}\tensor\mathcal{H}^N$, such that
\begin{equation}\label{defn:pseudopure}
    \rho = \epsilon\sigma + \frac{1-\epsilon}{MN}I,
\end{equation}
for $\sigma=\ketbra{\psi}{\psi}$ a pure state of dimension $MN$,
and $0< \epsilon \leq 1$. (The case $\epsilon = 0$ leads trivially to
$\fum = 0$.)
Without loss of generality, we assume $M\geq N$
in the following. Using the Schmidt decomposition~\cite{NC00}, we
can write $\ket{\psi}=\sum_{k=0}^{N-1}a_k\ket{k}_A\otimes\ket{k}_B$, where
$a_k$ are non-negative real numbers (Schmidt coefficients) with
$\sum_{k=0}^{N-1}a_k^2=1$, and $\set{\ket{k}_A}_{k=0}^{N-1}$ are
the elements of the Schmidt basis for subsystem A (analogously for B).
We first prove the following useful lemma.

\begin{lemma}\label{l:distinctAbsVal}
    Let $\rho$, acting on $\mathcal{H}^{M}\tensor\mathcal{H}^N$,
 be a pseudopure quantum state as defined in Equation (\ref{defn:pseudopure}).
Then, for any $k$ such that  $a_k \neq a_j$ for all $j\neq k$,
and for any unitary $U^B$, if $[\rho_B,U^B]=0$, then $\abs{U^B_{k,k}}=1$.
\end{lemma}\begin{proof}
    We first write $U^B=\sum_{ln}\bra{l}U^B\ket{n}\ketbra{l}{n}$, for
$\set{\ket{n}}_{n=0}^{N-1}$ the Schmidt basis for subsystem $B$. Then:
    \begin{equation}\label{eqn:zero_diag_commutator}
            [\rho_B,U^B]=\epsilon\sum_{ln}(a_{l}^2-a_n^2)
\bra{l}U^B\ket{n}\ketbra{l}{n}.
    \end{equation}
    If two Schmidt coefficients
 of $\ket{\psi}$ differ in value, it therefore follows that the corresponding entry in $U^B$ must be $0$ in order for $\rho_B$ and $U^B$ to commute. Thus,
for unique $a_k$, row $k$ and column $k$ of $U^B$ must be all zeroes, except for position ${U^B}_{k,k}$, for which $\abs{{U^B}_{k,k}}=1$, since $U^B$ is unitary.
\end{proof}

We now show the main result of this section. For the remainder of our discussion, let us denote the maximal Schmidt coefficient of $\ket{\psi}$ as
$a_{m}=\max_{k} a_k$.

\begin{theorem}\label{thm:pseudopure}
    Let $\rho$, acting on $\mathcal{H}^{M}\tensor\mathcal{H}^N$,
 be a pseudopure quantum state as defined in Equation
(\ref{defn:pseudopure}). Then,
    \begin{equation}\label{eqn:pseudoformula}
        \fum=
            \begin{cases}
                \epsilon & \text{if $a_{m}^2\leq\frac{1}{2}$,}\\
                2\epsilon \, a_{m}\sqrt{1-a_{m}^2} &\text{otherwise.}
            \end{cases}
    \end{equation}
\end{theorem}

\begin{figure}\centering
  \includegraphics[width=83mm]{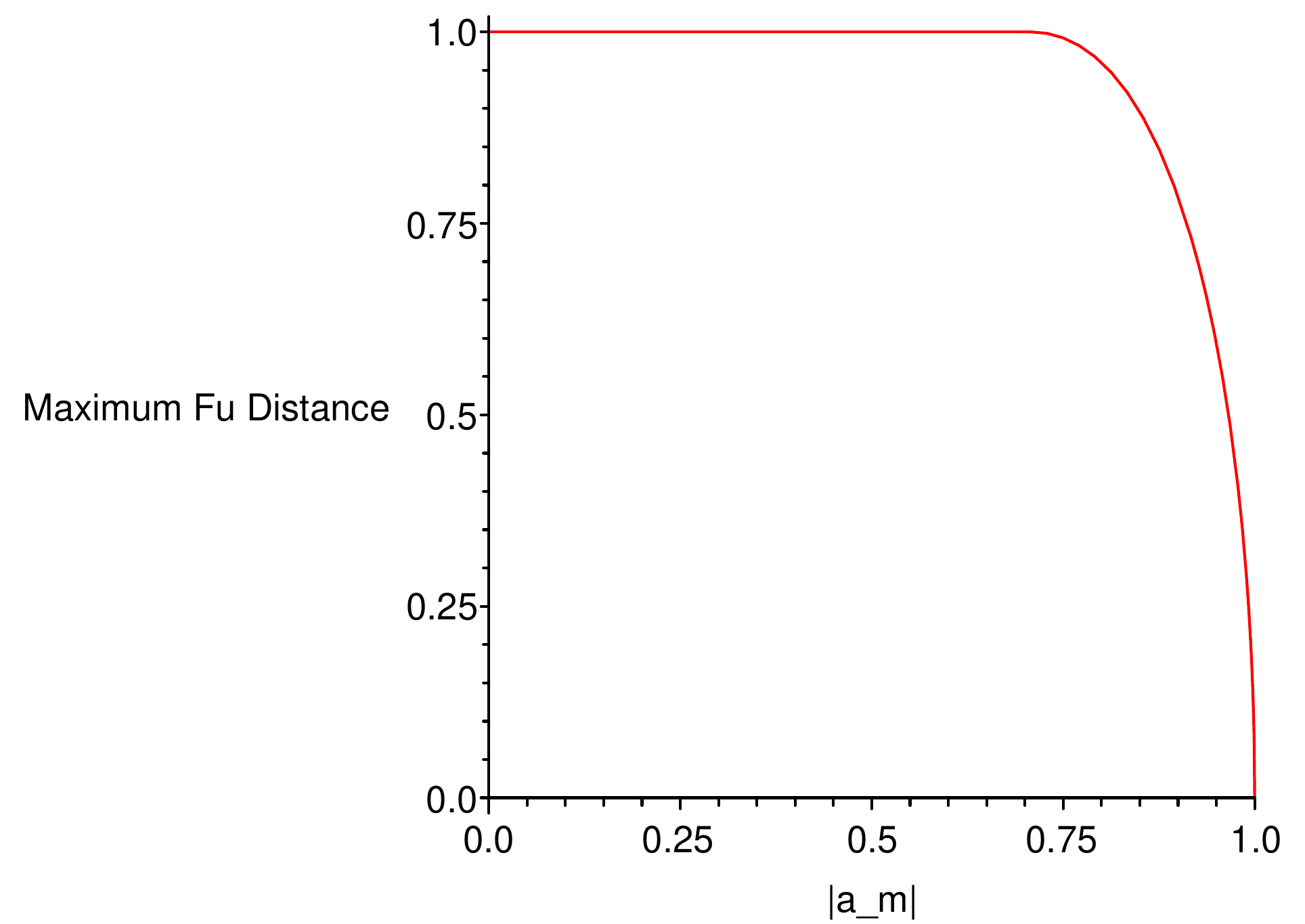}
  \caption{A plot of $\fum$ for pure states as a function of $a_{m}$,
as given by Equation~(\ref{eqn:pseudoformula}).
Note that only the region $a_m\geq 1/\sqrt{N}$ is accessible,
where $N$ is the smaller dimension of the two subsystems.
}\label{fig:pseudo}
\end{figure}

\begin{proof}
   Inserting  $\rho$ of Equation~(\ref{defn:pseudopure})
into Equation~(\ref{eqn:fudistancerewritten}) leads for arbitrary
$U^B$ (not necessarily cyclic), to
    \begin{equation}\label{eqn:simple_pseudo}
        \fu=\epsilon\sqrt{1-\left|\sum_{k=0}^{N-1}a_k^2\,
                             \bra{k}U^B\ket{k}\right|^2},
    \end{equation}
    from which it follows that $\fu$ depends only on the diagonal entries of $U^B$. Let us hence first assume that $U^B$ is a diagonal unitary matrix with eigenvalue $e^{i\theta_k}$ on row $k$ (for $\theta_k$ to be chosen), and subsequently show that choosing $U^B$ in this way is always optimal. By
Equation~(\ref{eqn:simple_pseudo}), we then find that
maximizing $\fu$ reduces to minimizing
$\abs{\sum_{k=0}^{N-1}a_k^2\, e^{i\theta_k}}$. Since $\rho_B$ is diagonal, any choice of $\theta_k$'s constitutes a commuting unitary operation $U^B$, and so this minimization problem has a simple geometric solution as follows.

    If $a_{m}^2\leq \sum_{k\neq m}a_k^2$ (or equivalently,
$a_{m}^2\leq 1/2$), then one can always construct a closed polygon using
 vectors of the lengths $a_k^2$ each exactly once. Hence,
$\min_{\set{\theta_k}_k}\abs{\sum_{k=0}^{N-1}a_k^2\, e^{i\theta_k}}=0$,
and so $\fum=\epsilon$. If, however, $a_{m}^2 > \sum_{k\neq m}a_k^2$, then
 no such polygon can be constructed, and the best minimization strategy is simply to set $\theta_{m}=0$ and $\theta_k=\pi$, for all $k\neq m$.
Substitution into our simplified expression for
Equation~(\ref{eqn:simple_pseudo}) and using the normalization
$\sum_{k=0}^{N-1}a_k^2=1$ then gives
$\fum=2\epsilon a_{m} \sqrt{1-a_{m}^2}$, as desired.

    Finally, to see that choosing $U^B$ diagonal is always optimal, note
that if $a_{m}^2\leq \sum_{k\neq m}a_k^2$, then our strategy for
diagonal $U^B$ achieves the maximum possible value for
Equation~(\ref{eqn:simple_pseudo}). If $a_{m}^2 > \sum_{k\neq m}a_k^2$,
on the other hand, clearly $a_{m}\neq a_k$ for all $k\neq m$, and so it follows from Lemma~\ref{l:distinctAbsVal} that $\abs{\bra{m}U^B\ket{m}}=1$ in
Equation~(\ref{eqn:simple_pseudo}). Thus, the best minimization strategy is again the same as that outlined for the diagonal case.
This concludes the proof.
\end{proof}
Let us point out some consequences of Theorem~\ref{thm:pseudopure}.
First, from Equation~(\ref{eqn:ccupperbound}), we know that a necessary
condition for  using $\fum$ to detect entanglement in $\rho$,
acting on $\mathcal{H}^{M}\tensor\mathcal{H}^N$, is $\fum>1/\sqrt{2}$.
Thus, from Theorem~\ref{thm:pseudopure} and straightforward calculation,
one finds that $\fum$ may be used to detect entanglement in
pseudopure states $\rho$ only if
\begin{equation}
    \frac{1}{2}\left(1-\sqrt{1-\frac{1}{2\epsilon^2}}\right)\leq
a_m^2
\leq \frac{1}{2}\left(1+\sqrt{1-\frac{1}{2\epsilon^2}}\right).
\end{equation}
For two-qubit pure states, this becomes an if and only if condition, and
the corresponding bound on $a_m$ simplifies to
$a_m\lesssim0.924$ (the lower bound in this case is implicitly given by $1/\sqrt{2}$, by definition of $a_m$). Second, for general dimensions, if one knows that $\ket{\psi}$ is a maximally
entangled state of the form
$\ket{\psi}=\frac{1}{\sqrt{D}}\sum_{k=0}^{D-1}\ket{kk}$ for $D=M=N$,
then it follows from Theorem~\ref{thm:pseudopure} that one can
always reliably detect the entanglement of the pseudopure
state $\rho$, since in this case $\rho$ is entangled if and only if
$\epsilon > 1/(D+1)$~\cite{PR00}. Third, for pure $\rho$, Theorem~\ref{thm:pseudopure} implies that one achieves $\fum=1$ as long as $a_{m}\leq1/\sqrt{2}$. Hence, $\rho$ need not  be maximally entangled in order to achieve a maximal shift. This surprising behavior is plotted in Figure~\ref{fig:pseudo}. Finally, it is clear from Theorem~\ref{thm:pseudopure} that $\fum>0$
for any entangled pseudopure $\rho$.

Is there any connection between the Fu distance and the concurrence?
This is the case for a
two-qubit pure state, i.e. $\ket{\psi}=a_0\ket{00}+a_1\ket{11}$ and
$\epsilon=1$. Then Equation~(\ref{eqn:pseudoformula}) reduces to
$\fum=2 a_0a_1=C(\rho)$, where $C(\rho)$ denotes the \emph{concurrence} of $\rho$~\cite{HW97}. Used by Wootters~\cite{W98} to derive an analytic formula for the entanglement of formation of two-qubit states, and an entanglement measure in its own right, the concurrence has a number of generalizations to higher dimensions~\cite{U00,RBCHM01,AVM01,W01,BDHHH02}, two of which have simple closed forms for the case of pure states, which we shall compare to
Equation~(\ref{eqn:pseudoformula}) of Theorem~\ref{thm:pseudopure} here. To do so, let $\rho$, acting on $\mathcal{H}^{M}\tensor\mathcal{H}^N$,
 be pure, and set $D=\min\set{M,N}$. Then, Rungta et al.~\cite{RBCHM01}
define the concurrence $C_R(\rho)$ for a pure state $\rho$ as
\begin{equation}
    C_R(\rho)=\sqrt{1-\trace(\rho_A^2)},
\end{equation}
where $0\leq C_R(\rho)\leq \sqrt{2(D-1)/D}$.
This expression can be rewritten as
\begin{equation}
    C_R(\rho)=\sqrt{2\left(1-\sum_{k=0}^{D-1}a_k^4\right)},
\end{equation}
from which it is clear that Equation~(\ref{eqn:pseudoformula}) does not reduce to (a normalized version of) $C_R(\rho)$. As a supporting example, consider $\ket{\psi}=\frac{1}{\sqrt{2}}\ket{00}+\frac{1}{2}\ket{11}+\frac{1}{2}\ket{22}$, for which $\fum=1$, but $C_R(\rho)\approx0.9682$, where we have normalized the latter by the maximum value possible for qutrits, $2/\sqrt{3}$.  Next, consider the generalization of Audenaert et al.~\cite{AVM01}, which states that for pure $\rho$, we have $C_A(\rho)=2a_m a_{m2}$, where $a_m$
and $a_{m2}$ are the first and second largest Schmidt coefficients in the Schmidt decomposition of $\ket{\psi}$. Again, it is clear that
Equation~(\ref{eqn:pseudoformula}) does not reduce to this definition $C_A(\rho)$ either. As an example, consider the maximally entangled two qutrit
state $\ket{\psi}=\frac{1}{\sqrt{3}}(\ket{00}+\ket{11}+\ket{22})$, for which $\fum=1$, but $C_A(\rho)=2/3$.

Note that in Figure~\ref{fig:pseudo} it is already evident that
for $D\geq 3$
several non-maximally entangled states lead to the same maximal
Fu distance, and therefore the Fu distance
can in general not be used to define an entanglement measure.

\subsection{Werner States}\label{sscn:werner}

We now turn our attention to bipartite Werner states $\rhow$ acting on
$\mathcal{H}^{D}\tensor\mathcal{H}^D$  with $D\geq 2$, for which we derive a closed formula for $\fuw$. Denoting as
$\{\ket{i}\}_{i=0}^{D-1}$ an arbitrary orthonormal basis for
$\mathcal{H}^{D}$, the Werner state $\rhow$ can be defined as follows~\cite{W89}:
\begin{eqnarray}
    P &=& \sum_{ij}\ketbra{i}{j}\otimes\ketbra{j}{i}\\
    P_{sym} &=& \frac{1}{2}(I_{D^2}+P)\\
    P_{as} &=& \frac{1}{2}(I_{D^2}-P)\\
    \rhow &=& p\frac{2}{D^2+D}P_{sym} +
    (1-p)\frac{2}{D^2-D}P_{as}, \label{eqn:wernerstatedef}
\end{eqnarray}
\noindent where $I_{D^2}$ is the $D^2$-dimensional identity matrix and
$0\leq p \leq 1$. The state $\rho$ is invariant under operation $U\tensor U$, for any choice of unitary $U$, and is entangled for $p<1/2$, and separable otherwise. Investigating in terms of Fu distance, we find the following result.

\begin{theorem}
    Let $\rhow$, acting on
$\mathcal{H}^{D}\tensor\mathcal{H}^D$, be a Werner state, as defined in Equation~(\ref{eqn:wernerstatedef}). Then
        \begin{equation}\label{eqn:wernerFuMax}
            \fuw=\frac{|2pD-D-1|}{D^2-1},
        \end{equation}
    obtained using any traceless $D\times D$ choice of unitary $U^B$.
\end{theorem}

\begin{proof}
    Consider substitution of $\rhow$ and arbitrary $U^B$ into
Equation~(\ref{eqn:fudistancerewritten}). Observing that $\trace(P)=D$,
    $\trace(P^2)=D^2$, and defining for convenience $\beta:=
    \trace(P(I\tensor U^B)P(I\tensor{U^B}^\adjoint))=\trace(U^B)\trace({U^B}^\adjoint)$, straightforward manipulation leads us to
        \begin{equation}\label{eqn:werner1}
            \operatorname{d}(\rhow,U^B) = \frac{\sqrt{(2pD-D-1)^2(D^2-\beta)}}{D(D^2-1)}.
        \end{equation}
    Examining the boundary and critical points of the first
    derivative of Equation~(\ref{eqn:werner1}) with respect to
$\beta$, we find that the two cases of interest are
    $\beta=-D$ and $\beta=0$. Note, however, that
    $\beta=-D$ implies $\trace(U^B)\trace({U^B}^\adjoint)=-D$, which
    is impossible, since $aa^\ast\geq0$ for all $a\in\complex$. Hence,
    the maximum Fu distance is achieved when $\beta=0$, implying that $U^B$ is
    traceless, giving the desired result.
\end{proof}

\begin{figure}\centering
  \includegraphics[width=54mm]{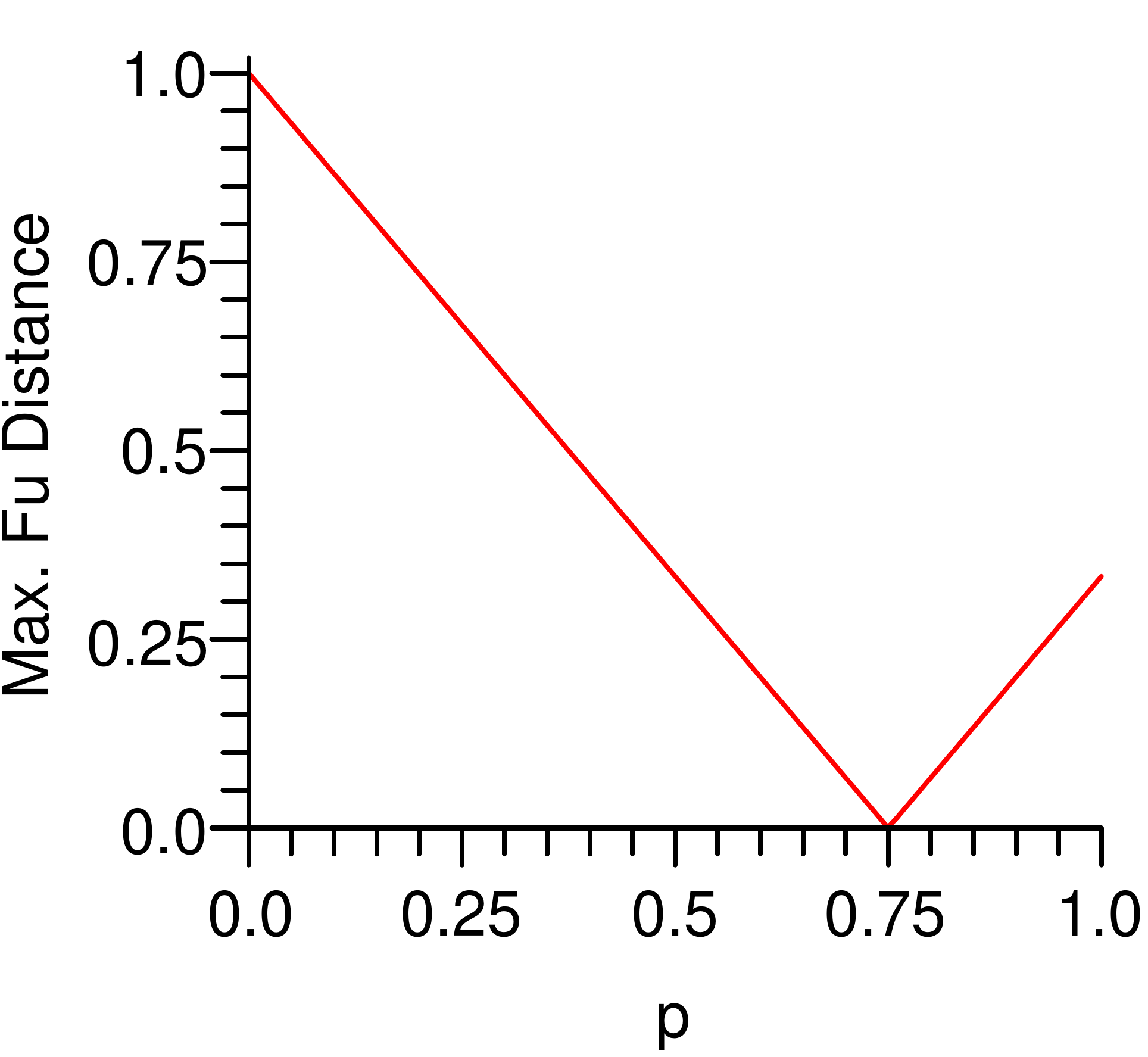}
  \includegraphics[width=54mm]{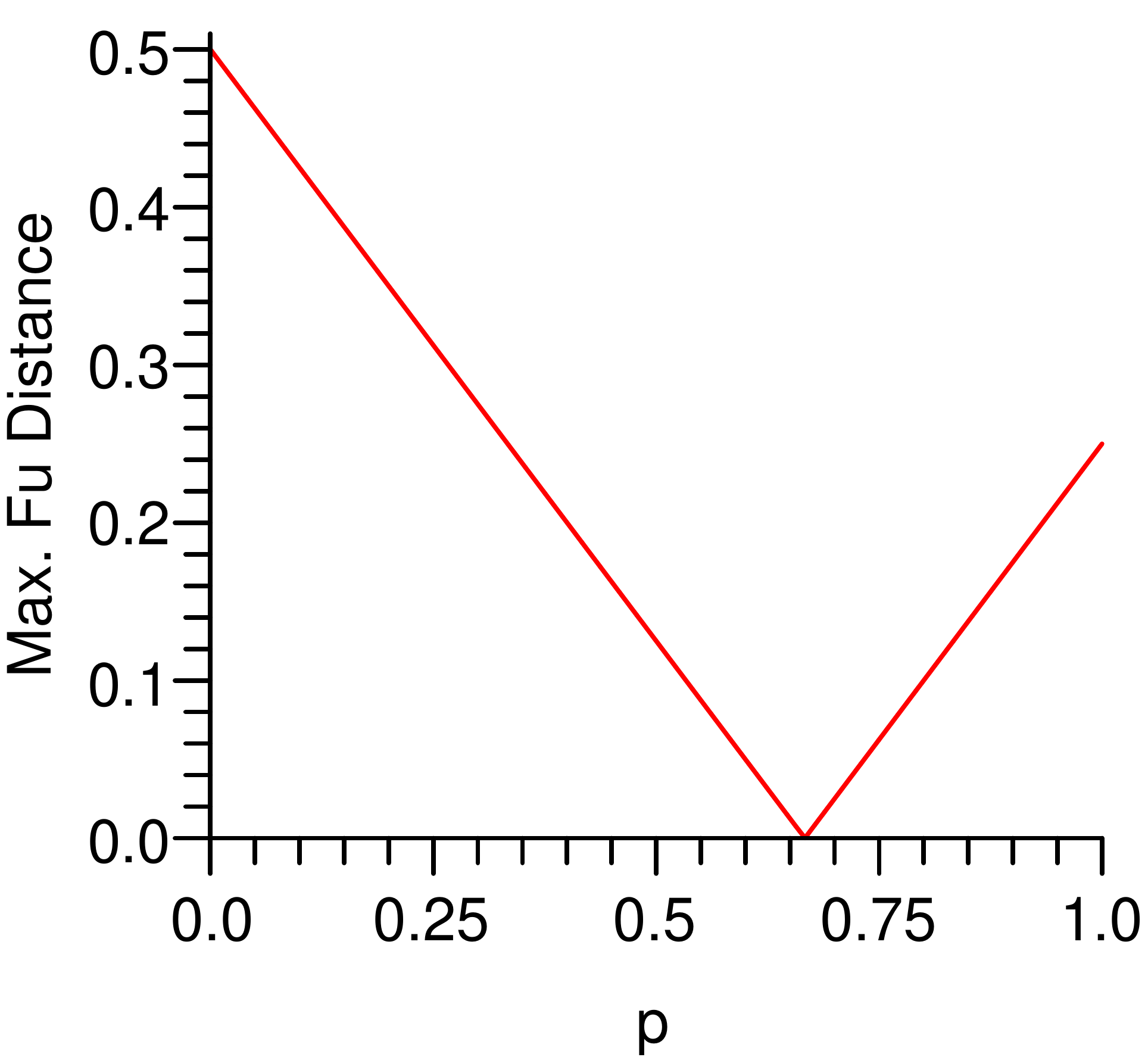}
  \includegraphics[width=54mm]{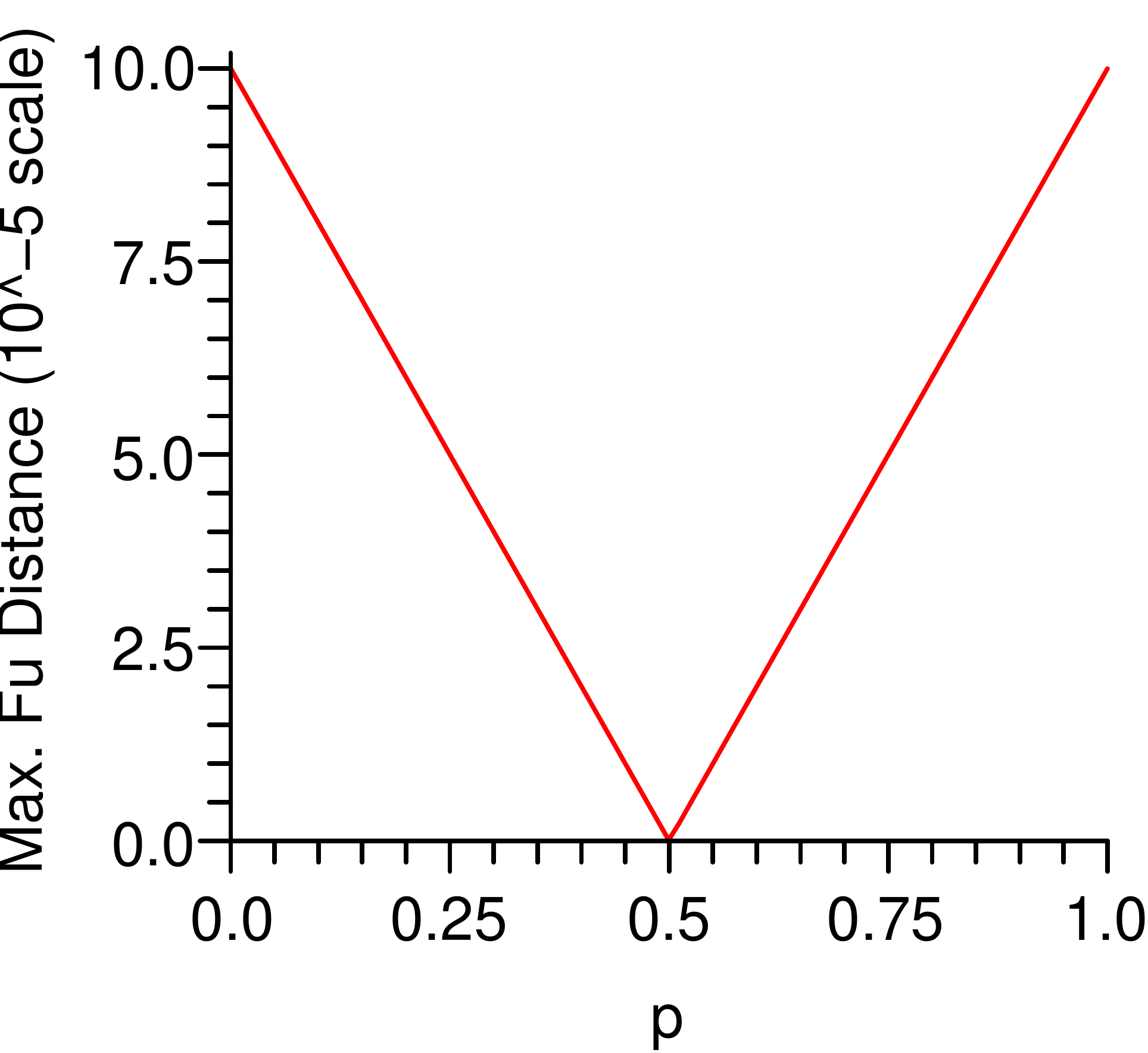}
  \caption{$\fuw$ plotted for $D=2,3,10000$,
  respectively, as given by Equation~(\ref{eqn:wernerFuMax}). Note the scale of $10^{-5}$ for the vertical axis for the case of $D=10000$.}\label{fig:1}
\end{figure}

We remark that for a two-qubit Werner state $\rhow$,
Equation~(\ref{eqn:wernerFuMax}) reduces to
Equation~(\ref{eqn:pseudoformula}), as required, since in this case $\rhow$ can be written as the pseudopure state
\begin{equation}
    \rhow =
|1-\frac{4}{3}p|\cdot \ketbra{\psi^-}{\psi^-}+(1-|1-\frac{4p}{3}|)\cdot
 \frac{I}{4}
\end{equation}
where $\ket{\psi^-}=1/\sqrt{2}(\ket{01}-\ket{10})$
and $0\leq p \leq 1$.

 We now direct the reader's attention to Figure~\ref{fig:1}, which
graphically depicts $\fuw$ for various dimensions $D$. For $D=2$,
we find from Equation~(\ref{eqn:ccupperbound}) that one can detect
entanglement in Werner states if $p\leq(3\sqrt{2}-3)/4\sqrt{2}\approx0.220$.
For arbitrary $D$, if we are promised that $\rhow$ is a Werner state, but
not given the value of $p$, then by Figure~\ref{fig:1}, attaining
$\fuw>1/(D+1)$ is sufficient to conclude that $\rhow$ is entangled,
seen by considering the case $p=1$ which leads to the maximal Fu distance
for a separable Werner state.
From Equation~(\ref{eqn:wernerFuMax}), it is also clear that
$\fuw>0$ for all entangled Werner states.

We remark that examining the critical points of the first derivative of
Equation~(\ref{eqn:wernerFuMax}) as $D\rightarrow\infty$ suggests that
there is a ``kink'' in the graph at $p=1/2$, which is precisely the boundary between entangled and separable Werner states. We also see that as
$D\rightarrow\infty$, $\fuw\rightarrow 0$, such that the possibility of distinguishing between classical and quantum correlations vanishes. To explain these phenomena (at least in a mathematical sense), we invoke the following theorem.

\begin{theorem}[\cite{DG}]\label{thm:fuBound}
    For any state $\rho$ acting on $\mathcal{H}^M\otimes\mathcal{H}^N$,
 $\fum\leq\sqrt{2\left[\trace(\rho^2)-\frac{1}{MN}\right]}$.
\end{theorem}

\noindent Intuitively, this means that the maximal Fu distance for a state is upper bounded by a dependence on the state's mixedness. In our case, straightforward calculation yields for the Werner state $\rho$:
\begin{equation}\label{eqn:wernermixedness}
    \trace(\rho^2)=p^2\frac{2}{D^2+D}+(1-p)^2\frac{2}{D^2-D}.
\end{equation}
The first derivative of this reveals that $\trace(\rho^2)$ has a minimum at
$\frac{1+1/D}{2}$. Thus, as $D\rightarrow\infty$, $\rho$ is most mixed at $p$ approaching $1/2$, explaining the first phenomenon above. Similarly, one finds
from Equation~(\ref{eqn:wernermixedness}) that $\trace(\rho^2)\rightarrow 0$
as $D\rightarrow\infty$, explaining the second phenomenon.

\section{Connections to the CHSH Inequality}\label{scn:CHSHconnections}

We now investigate connections between the CHSH inequality and the Fu distance for two-qubit systems. Our motivation stems from the following observation.
 From Equations~(\ref{eqn:pseudoformula}) and~(\ref{eqn:ccupperbound}), it is immediate that in order to use $\fum$ to detect entanglement in the
 two-qubit Werner state $\rhow=p\density{\psi^-}+\frac{1-p}{4}I$, where
$\ket{\psi^-}=\frac{1}{\sqrt{2}}(\ket{01}-\ket{10})$, we require $p>1/\sqrt{2}$. On the other hand, define the following quantity for any quantum
state $\rho$, acting on $\mathcal{H}^2\otimes\mathcal{H}^2$:
\begin{equation}\label{defn:M}
   M(\rho):=\tau_1(T^TT)+\tau_2(T^TT),
\end{equation}
where $T$ is the correlation matrix of $\rho$ from
Equation~(\ref{eqn:fano}), $T^T$ its transpose, and $\tau_1(T^TT)$ and
$\tau_2(T^TT)$ the first and second largest eigenvalues of $T^TT$, respectively. Then, $\rho$ can violate the CHSH inequality if and only if~\cite{HHH95}
\begin{equation}\label{eqn:MviolCHSH}
    M(\rho)>1.
\end{equation}
For  the Werner state
$\rhow$, one has $M(\rhow)=2p^2$, and so in order to detect entanglement in $\rhow$ using the CHSH inequality, one requires $p>1/\sqrt{2}$, which is the same bound obtained above for the Fu distance. Thus, we will pose and answer
the question of
whether there  is a connection between the ability to detect entanglement
via the CHSH inequality versus the Fu distance.

Our approach is as follows. We first show that, without loss of generality, one can take the correlation matrix $T$ of $\rho$ to be diagonal. We then derive a closed formula for $\fum$ for any two-qubit state $\rho$ with diagonal $T$. Using this formula, we  compare $\fum$ and $M(\rho)$.

To begin, we follow~\cite{HH96,HH96_2} and note that applying a unitary operation $U_1\tensor U_2$ to $\rho$ is the equivalent of applying orthogonal rotation matrices $O_1$ and $O_2$ to $\vec{r}^A$, $\vec{r}^B$, and $T$, such that:
\begin{equation}\label{eqn:su2so3}
    \vec{r}^A_f=O_1\vec{r}^A\quad\quad\quad\quad
    \vec{r}^B_f=O_2\vec{r}^B\quad\quad\quad\quad
    T_f=O_1TO_2^\adjoint
\end{equation}
Thus, given any $\rho$, we can find\footnote{This holds in the two-qubit case due to the existence of a surjective homomorphism from $SU(2)$ to $SO(3)$.} some $U_1\tensor U_2$ such that $\rho^\prime=U_1\tensor U_2\rho U_1^\adjoint\tensor U_2^\adjoint$ has diagonal $T$~\cite{HH96_2}. Further, by the following lemma, application of $U_1\tensor U_2$ to $\rho$ leaves $\fum$ invariant, implying we can assume without loss of generality that $T$ is diagonal, as desired.
\begin{lemma}\label{l:invarLocalOps}
    For any state $\rho$, acting on $\mathcal{H}^M\otimes\mathcal{H}^N$, $\fum$ is invariant under local unitary operations applied to $\rho$.
\end{lemma}
\begin{proof}
Let $U_1$ and $U_2$ be arbitrary unitary operations acting on subsystems $A$ and $B$, respectively. Then, straightforward manipulation of
 Equation~(\ref{eqn:fudistancerewritten}) yields $\operatorname{d}(U_1\tensor U_2\rho U_1^\adjoint\tensor U_2^\adjoint,U^B)=\operatorname{d}(\rho,U_2^\adjoint U^BU_2)$. It is easy to see that $[U_2\rho_BU_2^\adjoint,U^B]=0$ if and only if $[\rho_B,U_2^\adjoint U^BU_2]=0$. Thus, since the set of unitary matrices of fixed dimension forms a group, cycling through all possible choices of $U^B$ in $\operatorname{d}(\rho,U_2^\adjoint U^BU_2)$ gives the desired result.
\end{proof}

We can now derive a closed formula for $\fum$. Henceforth,
assuming that $T$ is diagonal, denote $\lambda_i:=T_{ii}$.

\begin{lemma}\label{l:maxFuDiagT}
    Given a quantum state $\rho$, acting on
$\mathcal{H}^2\otimes\mathcal{H}^2$, with diagonal correlation matrix $T$, we have
        \begin{equation}
\fum=\frac{1}{\sqrt{2}}\sqrt{\sum_{i=0}^2\lambda_i^2(1-n_i^2)} .
     \label{eqn:maxFuDiagT}
        \end{equation}
    \noindent
Here, if $\rho_B\neq I/2$, then $\vec{n}=\vec{r}_B/\enorm{\vec{r}_B}$, and otherwise $n_i=1$ for $\lambda_i=\min_k\lambda_k$ (with $\vec{n}= (n_0,n_1,n_2)^T$
and $\enorm{\vec{n}}=1$).
\end{lemma}

\begin{proof}
    Assume first that $\rho_B\neq I/2$. We shall manipulate
Equation~(\ref{eqn:fuT}) to achieve the claimed form. Specifically, let $U^B$ be a unitary operation corresponding to a rotation of angle $\theta\in[0,2\pi)$ about axis $\vec{n}=(n_0,n_1,n_2)^T$, with $\enorm{\vec{n}}=1$ ($\theta$ and $\vec{n}$ to be chosen as needed). We can characterize $T^f$ in terms of $T$ and $U^B$ by applying Equation~(\ref{eqn:su2so3}) for $U_1=I$ and $U_2=U^B$, and utilizing the following formula for $O_2$ in terms of $U^B$~\cite{G00}:
        \begin{eqnarray}
        O_2 &=& I + \sin\theta A + (1-\cos\theta)A^2\text{, where}\label{eqn:rot1}\\
        A &=& \left(
                \begin{array}{ccc}
                  0 & -n_2 & n_1 \\
                  n_2 & 0 & -n_0 \\
                  -n_1 & n_0 & 0 \\
                \end{array}
              \right).\nonumber
        \end{eqnarray}
    This simplifies Equation~(\ref{eqn:fuT}) to:
        \begin{equation}\label{eqn:twoqubitgeneral}
            \fu=\frac{1}{2}\sqrt{\sum_{i=0}^{2}\lambda_i^2(1-\cos\theta)(1-n_i^2)}.
        \end{equation}
    To choose $\vec{n}$, observe that demanding $[U^B,\rho_B]=0$  requires $U^B$ to induce a rotation about the Bloch vector of $\rho_B$ (unless $U^B=I$ or $\vec{r}_B= \vec{0}$). Thus, set $\vec{n}=\vec{r}_B/\enorm{\vec{r}_B}$.
Since $\enorm{\vec{n}}=1$, we have $(1-n_i^2)\geq0$ for all $i$, and so the expression above is maximized for $\cos\theta=-1$, or $\theta=\pi$, giving the desired result.

    Finally, if $\rho_B=I/2$, one can choose any axis of rotation
$\vec{n}$. By Equation~(\ref{eqn:twoqubitgeneral}), choosing $n_i=1$ for $\lambda_i=\min_k \lambda_k$ is  the optimal choice, as claimed.
\end{proof}

Lemma~\ref{l:maxFuDiagT} confirms that for any pure state $\ket{\psi}=a_0\ket{00}+a_1\ket{11}$ with $\abs{a_0}^2+\abs{a_1}^2=1$, $\operatorname{d}_{\operatorname{max}}(\psi)>0$ if and only if $\ket{\psi}$ violates the CHSH inequality~\cite{F06}. To see this, note that for $\rho=\ketbra{\psi}{\psi}$, $T$ is diagonal with  $T_{00}=2a_0a_1$, $T_{11}=-2a_0a_1$, $T_{22}=1$, and $\vec{r}_B=(0,0,a_0^2-a_1^2)^T$. Equation~(\ref{eqn:maxFuDiagT}) hence reduces to
Equation~(\ref{eqn:pseudoformula}), yielding $\operatorname{d}_{\operatorname{max}}(\psi)=2\abs{a_0a_1}$. The maximum violation of the CHSH inequality for a pure state is given by~\cite{G91}
\begin{equation}\label{eqn:BmaxPure}
 B_{\max}(\psi)=2\sqrt{1+4\abs{a_0a_1}^2}=2\sqrt{1+\operatorname{d}^2_{\operatorname{max}}(\psi)},
\end{equation}
where the inequality is violated if and only if $B_{\max}(\psi)>2$. The claim immediately follows. We now show the main results of this section.

\begin{theorem}\label{thm:onewayrelation}
    Given a quantum state $\rho$, acting on
$\mathcal{H}^2\otimes\mathcal{H}^2$, with diagonal correlation matrix $T$,
the implication
 $\fum>1/\sqrt{2} \Rightarrow M(\rho)>1$ holds. The converse is not true,
i.e. $M(\rho)>1 \not\Rightarrow\fum>1/\sqrt{2}$.
\end{theorem}
\begin{proof}
    Assume first that $\fum>1/\sqrt{2}$. Without loss of generality, let $\abs{\lambda_0}\geq\abs{\lambda_1}\geq\abs{\lambda_2}$, and let $\vec{n}=(\sqrt{\epsilon_0},\sqrt{\epsilon_1},\sqrt{\epsilon_2})^T$, where $\epsilon_0+\epsilon_1+\epsilon_2=1$. Then, substitution into Equation~(\ref{eqn:maxFuDiagT}) gives
    \begin{equation}\label{eqn:oneway1}
        (1-\epsilon_0)\lambda_0^2+(1-\epsilon_1)\lambda_1^2+(1-\epsilon_2)\lambda_2^2>1,
    \end{equation}
    from which it follows that $M(\rho)=\lambda_0^2+\lambda_1^2>1$, since setting $\epsilon_2=1$ can only increase the left hand side of
Equation~(\ref{eqn:oneway1}).

To show that the converse does not hold,
consider the following counterexample:
Given a pure state $\ket{\psi}=a_0\ket{00}+a_1\ket{11}$ (with
real coefficients $a_0$ and $a_1$, normalized via
$a_0^2+a_1^2=1$), for all $a_0\leq 0.3827$ or $a_0\geq 0.9239$, we have
$M(\rho)>1$, but $\fum\leq1/\sqrt{2}$.
\end{proof}

Theorem~\ref{thm:onewayrelation} implies that the Fu distance is generally a weaker entanglement criterion (at least in the two-qubit case) than the CHSH inequality. We next ask if  there are specific classes of two-qubit states for
which
the Fu distance is ``equivalent'' to the CHSH inequality, in the sense
that
$
    M(\rho)>1\Leftrightarrow \fum>\frac{1}{\sqrt{2}}
$
holds?
It turns out that this is indeed the case, as we will show now.

\begin{theorem}\label{thm:equivchshfu}
    Given a quantum state $\rho$, acting on
$\mathcal{H}^2\otimes\mathcal{H}^2$, with diagonal correlation matrix $T$
and its entries $\lambda_i$, with $i=0,1,2$, consider the following conditions:

    \begin{enumerate}
        \item $\lambda_i=\min_k\abs{\lambda_k}$, and $\abs{n_i}=1$, where $\vec{n}=\vec{r}_B/\enorm{\vec{r}_B}$, and $\vec{r}_B\neq(0,0,0)^T$.
        \item $\abs{\lambda_0}=\abs{\lambda_1}=\abs{\lambda_2}$.
        \item $\rho_B=I/2$.
    \end{enumerate}

    \noindent Then, $M(\rho)>1$ $\Leftrightarrow$ $\fum>\frac{1}{\sqrt{2}}$ if and only if one of the above conditions holds.
\end{theorem}

\begin{proof}
    We proceed case by case.

    \begin{enumerate}
        \item Suppose without loss of generality that $\lambda_2=\min_k\abs{\lambda_k}$, and $\vec{n}=(0,0,1)^T$. Then, Equation~(\ref{eqn:maxFuDiagT}) simplifies to
        \begin{equation}
                \fum=\frac{1}{\sqrt{2}}\sqrt{\lambda_0^2+\lambda_1^2}=\frac{1}{\sqrt{2}}\sqrt{M(\rho)},
        \end{equation}
        from which we have $M(\rho)>1$ $\Leftrightarrow$ $\fum>1/\sqrt{2}$.

        \item Suppose
        $\abs{\lambda_0}=\abs{\lambda_1}=\abs{\lambda_2}$. Then, since $\enorm{\vec{n}}=1$,
        Equation~(\ref{eqn:maxFuDiagT}) simplifies to:
        \begin{equation}
            \fum=\frac{1}{\sqrt{2}}\sqrt{\lambda_0^2(3-n_1^2-n_2^2-n_3^2)}=\frac{1}{\sqrt{2}}\sqrt{\lambda_0^2+\lambda_1^2}=\frac{1}{\sqrt{2}}\sqrt{M(\rho)},
        \end{equation}
        and we arrive at the same conclusion as in Case 1.

        \item Suppose $\rho_B=I/2$. Then by Lemma~\ref{l:maxFuDiagT}, it straightforwardly follows that we are reduced to to Case $1$.
    \end{enumerate}

    Finally, in order to show that the demonstrated equivalence holds if and only if one of these conditions
    hold, assume without loss of generality that
    $\abs{\lambda_0}\geq\abs{\lambda_1}\geq\abs{\lambda_2}$. Then,
    unless $\abs{\lambda_0}=\abs{\lambda_1}=\abs{\lambda_2}$ (Case 2), the only way to guarantee the equivalence is to have in
Equation~(\ref{eqn:maxFuDiagT})
the equalities $(1-r_0^2)=1$ and $(1-r_1^2)=1$, which
    implies that $\vec{r}_B=(0,0,\pm1)^T$. But such a choice of $\vec{r}_B$ can only
    correspond to a cyclic unitary operation if we have Case 1 or 3 above, as required.
\end{proof}

Thus, there exist certain classes of two-qubit states for which the CHSH inequality and $\fum$ are equally capable of detecting entanglement. Specifically, note that the Werner state $\rhow$ that we considered at the start of this section falls into such a class, since $\rho_B=\frac{I}{2}$ for $\rhow$,
 explaining the observed coincidence.

\section{Fu Distances for some Bound Entangled States}\label{scn:be}

Let us now investigate $\fube$ for three distinct constructions of bound entangled (BE) states of two qutrits in order to determine whether $\fube$ can be used to detect bound entanglement. Throughout this section, we denote the computational basis for qutrits as $\{\ket{0},\ket{1},\ket{2}\}$.

\subsection{P.~Horodecki Construction}\label{sscn:horodeckiBE1}
Denote by $P_\psi=\ketbra{\psi}{\psi}$ the projector onto a state $\kp$, and define~\cite{H97}:
\begin{eqnarray}
    Q &=& I\otimes I - (\sum_{k=0}^{2}P_{k}\otimes
    P_{k}) - P_{2}\otimes P_{0},\\
    \ket{\Psi} &=& \frac{1}{\sqrt{3}}(\ket{0}\otimes \ket{0} + \ket{1}\otimes \ket{1} + \ket{2}\otimes
    \ket{2}),\\
    \rhoent&=&\frac{3}{8}P_\Psi+\frac{1}{8}Q,\\
    \ket{\Phi_a} &=& \ket{2}\otimes (\sqrt{\frac{1+a}{2}}\ket{0} +
    \sqrt{\frac{1-a}{2}}\ket{2}),\\
    \rho_a &=& \frac{8a}{8a+1}\rhoent+\frac{1}{8a+1}P_{\Phi_a},
\end{eqnarray}
for $0\leq a \leq 1$. Note that $\rhoent$ is entangled, as its partial transpose has a
negative eigenvalue, and $P_{\Phi_a}$ is separable, since
$\ket{\Phi_a}$ is a product state. The state of interest, $\rho_a$, is bound entangled for $0<a<1$~\cite{H97}. Let us now determine $\fua$ in terms of $a$.

\begin{theorem}
    Given $\rho_a$, such that $0<a<1$, the maximal Fu distance is
 $\fuam=\frac{2\sqrt{2}a}{8a+1}$. It is obtained, for example,
by using any diagonal unitary matrix $U^B\in\complex^{3\times 3}$ with $U^B_{0,0}=-U^B_{1,1}=U^B_{2,2}$.
\end{theorem}
\begin{proof}
     Let $U^B$ be an arbitrary complex $3\times 3$ matrix, i.e.
    \begin{equation}
        U^B=\left(
            \begin{array}{ccc}
                u_1 & u_2 & u_3 \\
                u_4 & u_5 & u_6 \\
                u_7 & u_8 & u_9 \\
            \end{array}
        \right)\label{eqn:unknown3x3}.
    \end{equation}
    Defining $\gamma:=\sqrt{1-a^2}$, we have
    \begin{equation}
        [\rho_B,U^B] =\frac{1}{2(8a+1)}\left(
            \begin{array}{ccc}
            \gamma(u_7-u_3) & (1-a)u_2+\gamma u_8 & \gamma(u_9-u_1) \\
            u_4(a-1)-\gamma u_6 & 0 & u_6(a-1)-\gamma u_4 \\
            \gamma (u_1-u_9) & (1-a)u_8+\gamma u_2 & \gamma(u_3-u_7) \\
                                                         \end{array}
                                                       \right).
    \end{equation}
    Since $0<a<1$, it is easy to see that for $U^B$ to be cyclic it must therefore be of the form:
    \begin{equation}
        U^B= \left( \begin{array}{ccc}
                        u_1 & 0 & u_3 \\
                         0 & u_5 & 0 \\
                        u_3 & 0 & u_1 \\
                                  \end{array}
                                    \right).
    \end{equation}
    Inserting this into Equation~(\ref{eqn:fudistancerewritten})
 and enforcing unitary constraints on $U^B$ leads to
 (where $\operatorname{Re}(x)$ denotes the real part of $x\in\complex$)

    \begin{equation}\label{eqn:h1}
        \fua=\frac{a}{8a+1}\sqrt{6-2\left(|u_1|^2+2\operatorname{Re}(u_1^\ast u_5)\right)}.
    \end{equation}

To maximize $\fua$, we need to minimize
$|u_1|^2+2\operatorname{Re}(u_1^\ast u_5)$. To do so, set $u_5=-1$ and let $u_1=re^{i\theta}$ for $0\leq r \leq 1$, $\theta\in[0,2\pi)$. Then,
\begin{equation}
    |u_1|^2+2\operatorname{Re}(u_1^\ast u_5)=r^2-2r\cos(\theta),
\end{equation}
which  achieves a minimum at $\theta=0$ and $r=1$, or equivalently for $u_1=1$ and $u_5=-1$, yielding $\fuam=(2\sqrt{2}a)/(8a+1)$, as claimed. It is easy to see that any diagonal $U^B$ with entries $U^B_{0,0}=-U^B_{1,1}=U^B_{2,2}$ gives the same optimum value.
\end{proof}
The limiting value for $\fuam$ as $a$ approaches $1$ is
$\frac{2\sqrt{2}}{9}$.
By embedding a two-qubit state in a higher-dimensional space, one
finds that the value $d(\rho_{cc}) = 1/\sqrt{2}$ can be reached for a classically correlated state in any dimension.
Thus, we conclude that by using $\fuam$, one cannot detect entanglement in bound
entangled states of the above construction.

\subsection{Horodecki\superscript{$\tensor$3} Construction}

Consider now a second bipartite one-parameter qutrit bound entangled class of states due to Pawe\l, Micha\l, and Ryszard Horodecki~\cite{HHH99}. Define, for $2\leq \alpha \leq 5$:
\begin{eqnarray}
    \sigma_+&=&\frac{1}{3}(\ketbra{01}{01}+\ketbra{12}{12}+\ketbra{20}{20}),\\
    \sigma_-&=&\frac{1}{3}(\ketbra{10}{10}+\ketbra{21}{21}+\ketbra{02}{02}),\\
    \rho_\alpha&=&\frac{2}{7}\ketbra{\phi_+}{\phi_+}+\frac{\alpha}{7}\sigma_+
    +\frac{5-\alpha}{7}\sigma_-,
\end{eqnarray}
where $\ket{\phi_+}=\frac{1}{\sqrt{3}}(\ket{00}+\ket{11}+\ket{22})$. The state of interest, $\rho_\alpha$, is separable for $\alpha\in [2,3]$, bound entangled for $\alpha\in (3,4]$, and free
entangled for $\alpha\in(4,5]$. Determining an analytical form for $\fualm$ proves difficult, but if one is promised that the input state $\rho$ is of the form $\rho_\alpha$, but does not know $\alpha$, then choosing any $U^B$ with an all-zero diagonal (observing that $\trace_A(\rho_\alpha)=I/3$) gives
\begin{equation}
    \fual=\frac{1}{7}\sqrt{\alpha^2-5\alpha+9}\label{eqn:BEunitary2}.
\end{equation}
\begin{figure}\centering
  \includegraphics[width=80mm]{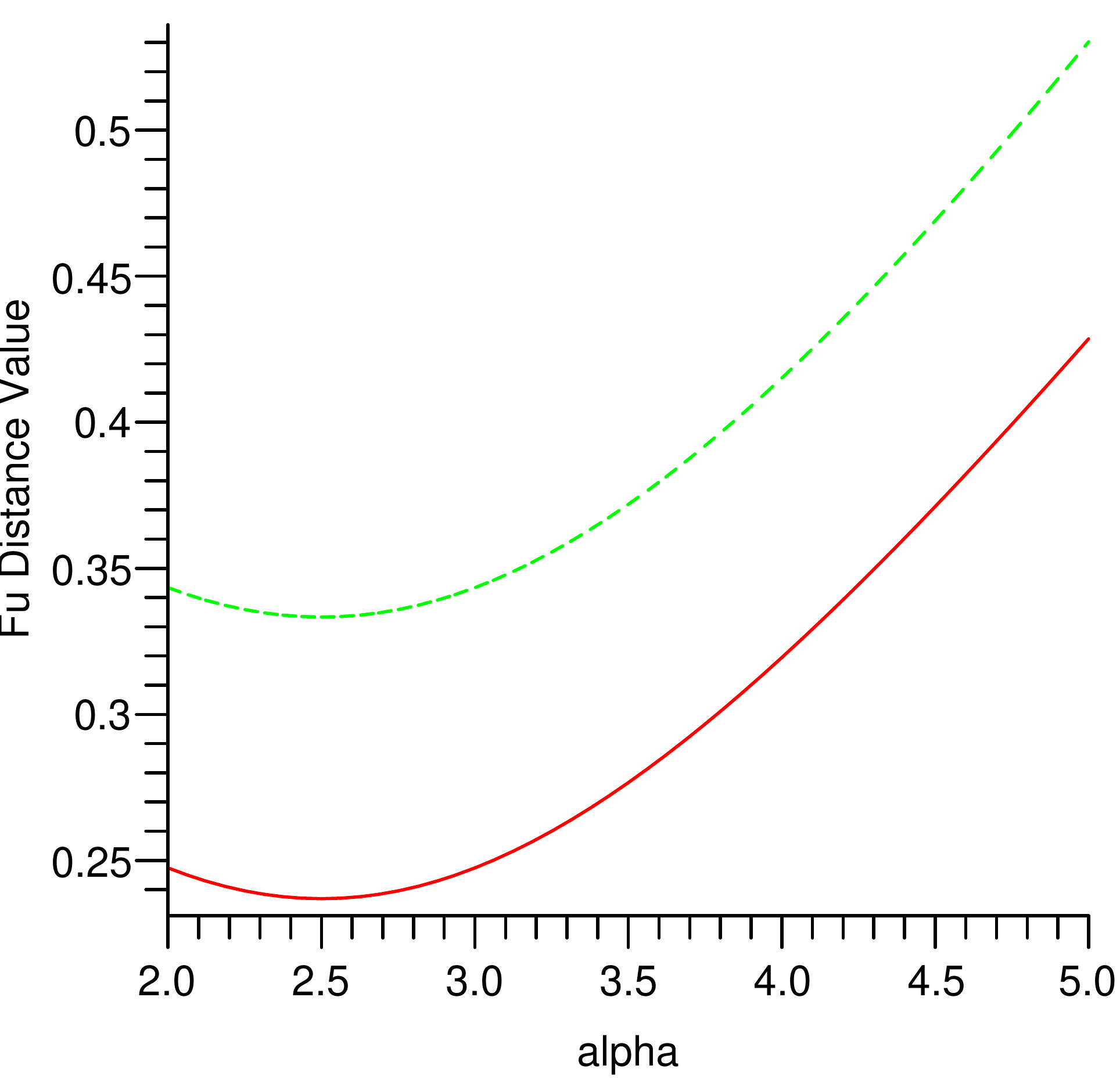}\\
  \caption{Plots of the Fu distance for the Horodecki et al. bound entangled state~\cite{HHH99}. The bottom solid line depicts $\fual$ for $U^B$ with an all-zero diagonal (see Equation~(\ref{eqn:BEunitary2})).
The top dashed line plots the bound on $\fualm$ given by Theorem~\ref{thm:fuBound} (see Equation~(\ref{eqn:limitbe})).}\label{fig:BE2}
\end{figure}
Straightforward calculation shows that the range of $\fual$ is disjoint for domains $\alpha\in [2,3]$,  $\alpha\in (3,4]$, and $\alpha\in(4,5]$, and so one can distinguish between all three cases using the Fu distance. This is illustrated in Figure~\ref{fig:BE2}. If one does not know that $\rho$ is of the form $\rho_\alpha$, on the other hand, we find via Theorem~\ref{thm:fuBound} that
\begin{equation}\label{eqn:limitbe}
    \fualm \leq \frac{2\sqrt{3\alpha^2-15\alpha+31}}{21},
\end{equation}
which is also plotted in Figure~\ref{fig:BE2}. For $\alpha\in(3,4]$, this gives $\fualm\leq2\sqrt{19}/21\approx0.415$, and for $\alpha\in(4,5]$, we have $\fualm\leq2\sqrt{31}/21\approx0.530$, and so in both cases we cannot detect bound entanglement using $\fum$.

\subsection{Unextendible Product Bases Construction}

We next consider the construction of Bennett et al.~\cite{BDMSST99}, which requires the following definition\footnote{This definition extends straightforwardly to the multipartite case~\cite{BDMSST99}.}.

\begin{defn}[Unextendible Product Basis (UPB)~\cite{BDMSST99}]
    Consider a bipartite quantum system in
$\mathcal{H}=\mathcal{H}_1\tensor\mathcal{H}_2$ with subsystems of arbitrary dimension. Define an incomplete orthogonal \emph{product basis} (PB) as a set $S$ of pure orthogonal product states spanning a proper subspace $\mathcal{H}_S$ of $\mathcal{H}$. Then an \emph{unextendible product basis} (UPB) is a PB whose complementary subspace $\mathcal{H}-\mathcal{H_S}$ contains no product state.
\end{defn}

Using a UPB, one can systematically construct BE states using the following theorem.

\begin{theorem}[Bennett et al.~\cite{BDMSST99}]\label{thm:UPBconstructionThm}
    Given UPB $\set{\ket{\psi}_i}_{i=0}^{n-1}$ in a Hilbert space of total dimension $D$, the following state is bound entangled:
    \begin{equation}\label{l:UPBBEstate}
        \rho=\frac{1}{D-n}\left(I-\sum_{k=0}^{n-1}\ketbra{\psi_k}{\psi_k}\right)
    \end{equation}
\end{theorem}

\noindent The UPB we shall use with Theorem~\ref{thm:UPBconstructionThm} is the following, given by~\cite{BDMSST99}:

\begin{eqnarray}
    \ket{\psi_0}&=&\frac{1}{\sqrt{2}}\ket{0}(\ket{0}-\ket{1}),\\
    \ket{\psi_1}&=&\frac{1}{\sqrt{2}}(\ket{0}-\ket{1})\ket{2},\\
    \ket{\psi_2}&=&\frac{1}{\sqrt{2}}\ket{2}(\ket{1}-\ket{2}),\\
    \ket{\psi_3}&=&\frac{1}{\sqrt{2}}(\ket{1}-\ket{2})\ket{0},\\
    \ket{\psi_4}&=&\frac{1}{3}(\ket{0}+\ket{1}+\ket{2})
(\ket{0}+\ket{1}+\ket{2}),\\
    \rho&=&\frac{1}{4}\left(I-\sum_{k=0}^{4}\ketbra{\psi_k}{\psi_k}\right).\label{eqn:UPB}
\end{eqnarray}

\noindent Using Theorem~\ref{thm:fuBound}, we find that for $\rho$,

\begin{equation}\label{eqn:UPBbound}
    \fum \leq \sqrt{\frac{2n}{D(D-n)}}.
\end{equation}

\noindent It thus follows that $\fum\leq 1/\sqrt{2}$ for any UPB for which $n\leq D^2/(D+4)$. Specifically, this rules out the possibility of detecting bound entanglement with the UPB chosen above, for which $D=9$ and $n=5$, yielding $\fum\leq\sqrt{10}/6\approx0.527$.

\section{Conclusions and Open Problems}\label{scn:conclusions}

We have investigated locally noneffective unitary operations in
connection with the detection of quantum entanglement. Specifically, we have derived and discussed closed formulas for $\fum$
 (the maximal distance between the original state $\rho$ and the state after
a locally noneffective unitary operation),
 for the bipartite cases of (pseudo)pure quantum states, Werner states, and two-qubit states.  The first of these reveals the existence of non-maximally entangled states capable of achieving a maximal global shift.
 Thus, no entanglement measure based on $\fum$ can be defined,
and similarities to anomalies seen in non-locality measures are revealed.
 Since $\fum$ is neither an entanglement measure, nor a non-locality measure, yet as demonstrated here possesses clear connections to the CHSH inequality in the two-qubit case, it would be interesting to have a better intuitive understanding of the correlations (both classical and quantum) that $\fum$ is quantifying, and if and how the anomalies mentioned above are related to those seen in non-locality measures.

There are a number of questions which remain open. First, despite the fact that our formula for $\fum$ for Werner states demonstrates diminishing distinguishability between classical and quantum correlations in Werner states as the dimension grows, it remains for a tight upper bound on $\fum$ to be found for classically correlated states of total dimension $D>4$ in order to conclusively state the efficacy of $\fum$ as an entanglement detection criteria. Second, although we have demonstrated that any entangled pseudopure or Werner state achieves $\fum>0$, it is still not known whether this holds for all entangled bipartite states. Third, it would be of interest to determine whether a closed formula for $\fum$ can be derived for mixed states in general, the existence of which would not contradict known hardness results for the quantum separability problem~\cite{G03}. Finally, as mentioned briefly in Section~\ref{scn:introduction}, the principle behind the Fu distance is implicitly applied in superdense coding, and we would be curious to know  whether there exist any other applications in quantum computing and information.

\section{Acknowledgements}

We would like to thank Richard Cleve, Andr\'{e} M\'{e}thot, Frank M\"{u}nchow,
and Marco Piani for helpful and intriguing
discussions regarding the Fu distance
and non-locality. This work was partially supported by Canada's NSERC, CIAR and MITACS, as well as the EU
Integrated Project SCALA.

\bibliographystyle{unsrt}
\bibliography{Bibliography}

\end{document}